\documentclass[sigconf]{acmart}
\usepackage{url}
\usepackage{balance}
\newcommand{\squishlist}{
 \begin{list}{$\bullet$}
  { \setlength{\itemsep}{0pt}
     \setlength{\parsep}{1pt}
     \setlength{\topsep}{1pt}
     \setlength{\partopsep}{0pt}
     \setlength{\leftmargin}{1.5em}
     \setlength{\labelwidth}{1em}
     \setlength{\labelsep}{0.5em} } }
 \newcommand{\squishend}{\end{list}}

\AtBeginDocument{%
  \providecommand\BibTeX{{%
    \normalfont B\kern-0.5em{\scshape i\kern-0.25em b}\kern-0.8em\TeX}}}

\copyrightyear{2020} 
\acmYear{2020} 
\setcopyright{acmcopyright}
\acmConference[CHIIR '20]{2020 Conference on Human Information Interaction and Retrieval}{March 14--18, 2020}{Vancouver, BC, Canada}
\acmBooktitle{2020 Conference on Human Information Interaction and Retrieval (CHIIR '20), March 14--18, 2020, Vancouver, BC, Canada}
\acmPrice{15.00}
\acmDOI{10.1145/3343413.3378011}
\acmISBN{978-1-4503-6892-6/20/03}

\begin{document}

\title{Personalized Entity Search by Sparse 
and Scrutable User Profiles}

\author{Ghazaleh H. Torbati}
\email{ghazaleh@mpi-inf.mpg.de}
\orcid{TODO?}
\affiliation{%
 \institution{Max Planck Institute for Informatics}
 \city{Saarbrücken}
 \country{Germany}
}
\author{Andrew Yates}
\email{ayates@mpi-inf.mpg.de}
\affiliation{%
 \institution{Max Planck Institute for Informatics}
 \city{Saarbrücken}
 \country{Germany}
}
\author{Gerhard Weikum}
\email{weikum@mpi-inf.mpg.de}
\affiliation{%
 \institution{Max Planck Institute for Informatics}
 \city{Saarbrücken}
 \country{Germany}
}

\begin{abstract}
Prior work on personalizing web search results has focused on considering query-and-click logs to capture users' individual interests. For product search, extensive user histories about purchases and ratings have been exploited. However, for general entity search, such as for books on specific topics or travel destinations with certain features, personalization is largely underexplored. In this paper, we address personalization of book search, as an exemplary case of entity search, by exploiting  sparse user profiles obtained through online questionnaires. We devise and compare a variety of re-ranking methods based on language models or neural learning.
Our experiments show that even very sparse information about individuals can enhance the effectiveness of the search results.

\end{abstract}



\maketitle
\pagestyle{plain}
\vspace{-0.2cm}
\section{Introduction}

\noindent{\bf Motivation and Problem:}
Personalization to improve web search result ranking has been 
a long-standing theme in information retrieval
\cite{DBLP:conf/sigir/TeevanDH05,DBLP:reference/db/WenDS18}.
With 
the increasing availability 
of 
individual users'
online traces and derived traits, 
personalization is again gaining importance
for chatbots, recommender systems, product search, and more.
\cite{DBLP:conf/ictir/BalogK19} has formulated a vision and
research agenda for constructing and leveraging 
{\em personal knowledge graphs (PKG's)} in such settings.
In this paper, we investigate the role of 
PKG's
for topical entity search, with the challenging case that the
only per-user knowledge is a sparse profile obtained from a short
questionnaire. 
In contrast to the ``data-hungry''
approaches of prior works, we focus on
this ``minimal PKG'' case to strengthen
the user's ability to understand and
control her user profile, similar to
what major search engines offer for controlling
the personalization of ads
(e.g., {\small\url{adssettings.google.com}}).
Note that our case is more ambitious, though:
minimal-PKG profiles aim to capture the bare necessities, whereas ads controls often
comprise a hundred or more tags for the same user.
The fewer traits the profile contains and
the more explicit they are (as opposed to
learned latent models), the more scrutable
and actionable the personalization model
becomes from a user perspective.

\vspace{0.1cm}
\noindent{\bf State of the Art and its Limitations:}
The most 
important
line of exploiting user information for 
general web search
is based on {\em query-and-click logs} (e.g., \cite{DBLP:conf/sigir/TeevanDH05,DBLP:journals/ftir/Silvestri10}).
This helps in interpreting user interests and
intents for ambiguous queries
as well as for identifying salient
pages for popular queries,
and for 
suggestions for query auto-completion
(e.g., \cite{DBLP:conf/sigir/Shokouhi13}).
In all this, cues about the user's location and daytime 
are a major asset, too
(e.g., \cite{DBLP:conf/sigir/BennettRWY11}).

{\em Recommender systems} have incorporated
personalization as well,
for ads, products and other contents (e.g., \cite{DBLP:reference/sp/2015rsh,DBLP:journals/tmis/Gomez-UribeH16,DBLP:journals/internet/SmithL17}).
Here, structured data 
is leveraged, 
most notably, purchases or ratings of products,
likes of news, YouYube videos, Instagram photos, etc. 
This field has recently paid attention to 
{\em scrutable recommendations} that are comprehensible
by end-users and pinpoint the specific data that
explains how the recommended item was
computed \cite{DBLP:journals/corr/abs-1804-11192,DBLP:conf/sigir/BalogRA19,DBLP:conf/recsys/PeiZZSLSWJGOP19}.
However, these approaches are at least
as ``data-hungry'' as the search engines,
and require extensive user-specific data.

{\em Entity search} about people, products or events has received
great attention and has been incorporated into major
search engines (see, e.g.,
\cite{DBLP:journals/ftir/BastBH16,DBLP:series/irs/Balog18} and further references there). 
This methodology leverages large knowledge graphs to infer the focus
of the query and/or return crisp entities as answers.
However, except for special cases such as music recommendation
\cite{Carterette2019}
and consumer product search \cite{ai2017learning},
there is hardly any work on {\em personalized} entity search
with individual user traits.

\vspace{0.1cm}
\noindent{\bf Approach:}
This paper explores the direction of personalized entity search, relying solely
on a ``minimal-PKG'' user profile for
scrutability.
The requirement is that users can fully
understand and control  the information that
drives the personalization (e.g., modify or
revoke pieces of a profile),
and that this is as sparse as possible
while still giving benefits. 
We consider online {\em questionnaires} 
as 
a source of sparse and scrutable user profiling.

\begin{figure}[ht!]
  \centering
     \includegraphics[width=0.8\linewidth]{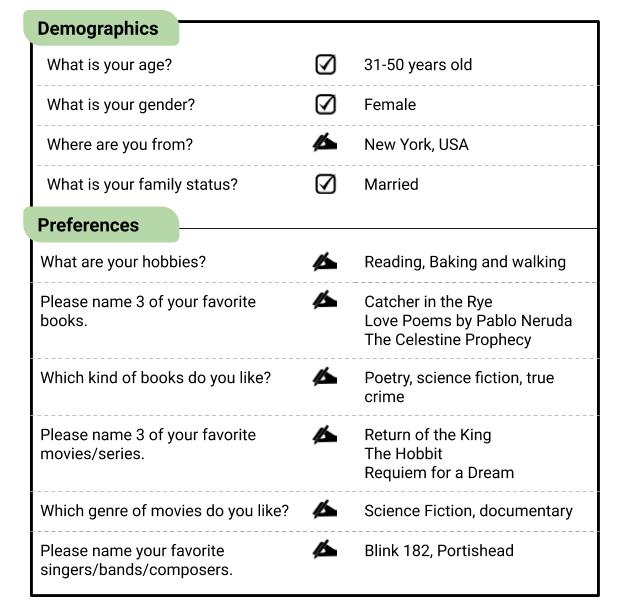}
\vspace{-0.15cm}
\caption{Sparse user profile from questionnaire.}
  \label{fig:pkb}
\end{figure}

\vspace{0.1cm}
\noindent{\bf Example:}
Figure \ref{fig:pkb} shows an excerpt of a
questionnaire, obtained by
hiring crowd workers at Amazon MTurk.
It captures demographic attributes
(age, gender, location etc.),
personal tastes regarding
books, movies and music, 
and hobbies -- all entered as
free-form text (as opposed to guiding users through menus
which may create bias).
The questionnaire has only 10 questions, and users spent 
7 minutes, on average, to fill in their answers.
\begin{figure}[ht!]
  \centering
  \includegraphics[width=0.8\linewidth]{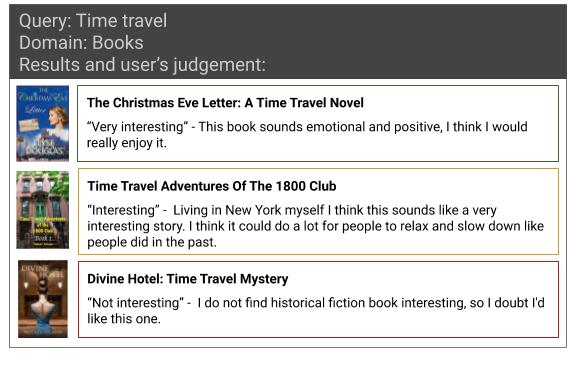}
\vspace{-0.15cm}
  \caption{Examples of entity search results for the query "time travel" and the user's judgements and justifications.}
  \label{fig:user_judgements}
\end{figure}
For entity search, we consider medium-grained query topics about
books: finer-grained than merely specifying a genre of interest
(e.g., ``history'' or ``science fiction''), but not as specific
as aiming for a single entity
as an answer. %
Examples of our queries are:
``historical romance'', 
``sci fi noir'', and
``time travel''. 

Consider the example query ``time travel''
on the domain of books, executed for entity pages returned from
{\small\url{goodreads.com}}. The results were personalized for
the participants of our MTurk study, including the user
whose profile is shown in Figure \ref{fig:pkb}.
Figure \ref{fig:user_judgements} shows three query results 
and the user's judgments along with her justification sentences. 
As shown in Figure \ref{fig:pkb}, she is from ``New York'',
and one of her favorite books  is  ``Love poems from Pablo Neruda''.
Personalized re-ranking over the top-100 answers correctly 
inferred that she would find an emotional book ``very interesting'' 
and a book with a New-York-based storyline ``interesting''.

\vspace{0.1cm}
\noindent{\bf Contribution:}
Our methodology focuses on ranking entities as candidate answers,
based on the sparse profiles from questionnaires.
We cast this background knowledge into user-specific language models or queries,
but treat entities (e.g., favorite books, movies or
singers/bands) in the user answers different from
free-form text (e.g., talking about hobbies).
To this end, we incorporate entity descriptions from general
knowledge graphs, and we also consider word embeddings (from word2vec)
as an additional input.
This way, we derive a suite of {\em re-ranking methods}
for pools of candidate answers obtained by running a standard
search engine on the book community portal {\small\url{goodreads.com}},
which has both content summaries for
millions of books and comments from nearly 100 million users.

The key hypothesis that we test in this study is that
even a very small PKG about a user can
improve the quality of search answers as assessed by the user herself.
To this end, we hired the same MTurk workers for judgements of
interestingness who contributed their questionnaire answers. In total, we evaluated 115 user-query pairs
from 33 distinct users. 

Our experiments compare a variety of re-ranking methods,
with different 
degrees of incorporating
sparse user profiles.
The results are preliminary, as the study is %
limited in scale
and scope. Nevertheless, our findings indicate that even
sparse profiles yield statistically significant benefits over not personalizing at all.

\vspace{-0.15cm}
\balance
\section{Related Work}
Prior works 
covered two major dimensions
\cite{Ghorab2013}:
\squishlist
\item[1.] Creating user models from explicit
signals like queries, clicks, likes, social links, etc. \cite{agichtein2006learning} or/and rich contents like
email histories or desktop data
\cite{DBLP:conf/sigir/ChiritaFN07,kuzi2017query}.
\item[2.] Leveraging this background knowledge for
answer ranking, query expansion, and auto-completion suggestion 
\cite{DBLP:conf/wsdm/MatthijsR11,DBLP:conf/sigir/Shokouhi13,DBLP:conf/sigir/CaiR16}.
\squishend

On the first dimension,
\cite{DBLP:conf/sigir/TeevanDH05} pioneered
the analysis of user interests and activities
reflected in query, click and mail histories,
and possibly even 
other online contents written or read by a user.
\cite{DBLP:conf/cikm/ShenTZ05} focused on short-term
context, like browser sessions, to infer the user's interest 
and personalize interactive search.
Numerous follow-up works addressed the analysis and usage of 
query-and-click logs and browsing sessions. 
To learn from this kind of expressive but highly noisy data, 
\cite{agichtein2006learning} introduced predictive models with learning-to-rank features, whereas \cite{DBLP:conf/sigir/ChiritaNPK05} and
\cite{Stamou2009} explored the use of similarity signals from taxonomies and ontologies.
\cite{DBLP:conf/ecir/YangGSMSMC16}
learned models of user interests for
proactive ``zero-query'' search.
\cite{DBLP:conf/sigir/BennettRWY11} studied the important role
of user location.

On the second dimension,
prior works explored personalization for
ranking as well as query expansion and query
suggestions.
For {\em personalized ranking}, \cite{DBLP:conf/wsdm/SontagCBWDB12}
developed methods for incorporating user-specific priors
into language models.
The interplay of a user's long-term behavior and short-term context
for personalized ranking
has been investigated in
\cite{DBLP:conf/sigir/BennettRWY11,DBLP:conf/sigir/BennettWCDBBC12}.
\cite{DBLP:conf/sigir/TeevanDL08} and
\cite{DBLP:conf/ictir/BennettSC15} addressed the issue
of selective personalization: when to incorporate user profiles.

Another line of research addresses {\em query expansion} 
for personalization. 
\cite{DBLP:journals/tist/BiancalanaGMS13a,zhou2017query} utilize folksonomy data, like
user-provided tags in social bookmarking communities,
as a source for expanding a user's queries.
\cite{kuzi2017query} personalizes email search 
via word embeddings learned from 
email histories. 
\cite{DBLP:conf/sigir/ChiritaFN07} proposes
methods for harnessing a user's
desktop files (incl. email).
The viability of all these methods relies on the availability 
of large collections of
user data.

The same assumption holds for prior work
on {\em query auto-completion} 
\cite{DBLP:conf/sigir/Shokouhi13,DBLP:conf/sigir/CaiR16},
perhaps the most successful line of 
personalization in major search engines.
The underlying user data ranges from long-term
query-and-click histories to browser histories
to email contents, in addition to location
and daytime as short-term context.

For {\em entity search},
to the best of our knowledge, 
prior work on personalization
is scarce.
CLEF had a series of competitions on book recommendations
\cite{CLEF16_10.1007/978-3-319-44564-9_29}, but this relied on posts, tags, reviews and ratings by 
many users in the
LibraryThing community and the Amazon shop.
The closest to our work is \cite{Ai:2017:LHE:3077136.3080813}
on personalized product search. 
It is based on learning embeddings for users and items in 
the same semantic space, by leveraging user-written reviews on item pages. However, such rich data about individual users
is not easily available for general entity search.

\vspace{-0.1cm}
\section{Methodology}

This section discusses how we cast user input from questionnaires into
user profiles, and how these 
are
incorporated into different kinds of rankers:
language models, BM25, and neural methods.
\vspace{-0.2cm}
\subsection{User Profiles}\label{sec:user_profile}
In contrast to prior works 
based
on extensive logs of user queries, clicks and other activities,
we focus on sparse and concise models of
user-specific interests and tastes. 
To this end, we designed a small questionnaire and hired
crowd workers at MTurk to fill in their answers.
As shown in Figure \ref{fig:pkb}, these profiles cover
basic demographics and personal attributes like
hobbies, favorite books and book genres, favorite movies
and movie genres, and favorite singers or music bands. 
The advantage of this ``minimal-PKG'' approach
is that such a profile is easily comprehensible by
the respective user, and easy to control as the user
may want to change it when her interests evolve or
she becomes concerned about privacy.
This kind of scrutability and controllability is impossible
with a huge log and latent models derived from massive
user data.

As most fields in the questionnaire are free-form text,
we treat the profiles as text documents,
like a statistical language model, or a bag-of-words model,
or a term-sequence model, depending on which ranking method
is adopted.

\vspace{0.05cm}
\noindent{\bf Incorporating Entity Descriptions:}
The user profiles include named entities like favorite books, movies
and musicians.
To take advantage of the sparse data to its fullest potential, we 
employ named entity disambiguation to link the entity mentions
to their respective entries in the Yago knowledge base
and in Wikipedia. Our implementation uses the AIDA
tool \cite{DBLP:conf/emnlp/HoffartYBFPSTTW11}; 
for experiments we manually checked (and corrected
a few of) the computed links to eliminate a potential
source of erroneous drift.
For the user models (language model or other),
we did not include the entity names 
themselves to avoid
overfitting to specific entities.
Instead, we incorporated the descriptions of these entities
from the knowledge base and Wikipedia.
For the test case of books, we selected the first paragraph
of each book's Wikipedia article.
Typically, this gives summary information about 
the
book's story.
\vspace{-0.2cm}
\subsection{Ranking Methods}\label{sec:method}
Given a pool $M$ of non-personalized results for a query $q$ and a user $u$, we want to re-rank the results such that 
the ranking reflects the user’s individual interests. 
All ranking methods treat query answers as text documents
about specific entities. In our experiments, these are
pages from {\small\url{goodreads.com}}, each featuring a
single book -- typically in the form of a content summary
and user comments.
\vspace{-0.1cm}
\subsubsection{\textbf{Statistical Language Models}}\label{methods_1}
This method 
adopts a query-likelihood model where the score of 
document $d \in M$ for query $q$ is proportional to the
Kullback-Leiber divergence between the language models of $q$ and $d$.
As we want to capture the interestingness of a document for
a user, rather than general relevance for the query, we
incorporate the user profile as another language model.
We use a mixture model over the
estimated language models $\theta_q$, $\theta_d$, $\theta_u$ for the query, document and user, respectively,
and and a general background model $\theta_C$ (based on
ClueWeb09) for smoothing.

\vspace{0.1cm}
\noindent{\bf Incorporating Word Embeddings:}
Optionally, we integrate word embeddings; our implementation
uses pre-computed word2vec vectors.
The language model is augmented by a translation model,
largely following \cite{DBLP:conf/cikm/KuziSK16}.
Using the cosine between word-embedding vectors as 
term-term similarity $sim$.
In the following ranking equation,
$div$ is the Kullback–Leibler divergence, 
$p(w|u) = \frac{sim(w,u)}{\sum_{u'\in V_d}{sim(u',u)}}$
are the probabilities for the word-word relatedness model,
$V_q$ and $V_u$ denote the query and user vocabularies, $p(w|\theta)$ 
are the 
estimated 
probabilities
of word $w$ in each language model $\theta$, $\mu$ is the Dirichlet smoothing parameter, and $\lambda$ determines the 
relative weight of the query and user models. 
\vspace{-0.1cm}
\begin{equation}
\begin{split}
rank(q,d) \propto \lambda div(\theta_q\| \theta_{d_\mu}) + (1-\lambda) div(\theta_u\| \theta_{d_\mu}) = \\
\lambda \sum_{w\in V_q}{p(w|\theta_q)\log\frac{p(w|\theta_q)}{\frac{ \sum_{u\in V_d}{p(w|u) p(u|\theta_d)} + \mu p(w|\theta_C)}{|d| + \mu}}} +\\
(1-\lambda) \sum_{w\in V_u}{p(w|\theta_u)\log\frac{p(w|\theta_u)}{\frac{\sum_{u\in V_d}{p(w|u) p(u|\theta_d)} + \mu p(w|\theta_C)}{|d| + \mu}}}
\end{split}
\label{eq2}
\end{equation}
\vspace{-0.2cm}
\subsubsection{\textbf{BM25 Ranking}}\label{methods_2}
To personalize the BM25 scoring, we use query expansion,
treating the user profile as a bag-of-words.
As we use it only for re-ranking a pool of candidate results,
the query is fixed and the user profile becomes the actual query.
When incorporating entity descriptions, they are
included in this query.
\vspace{-0.5cm}
\subsubsection{\textbf{Neural Ranking}}\label{methods_3}
We adopted the state-of-the-art methods DRMM \cite{DBLP:conf/cikm/GuoFAC16}
and PACCR \cite{DBLP:conf/wsdm/HuiYBM18} as representatives for neural ranking.
Analogously to BM25, we focus on re-ranking
and thus treat the user profile as the query,
as a term sequence where each term is represented
as an embedding vector using word2vec.

\begin{table*}[th!]
\centering
\small
\begin{tabular}{@{}l ccc   ccc   ccc@{}}
\toprule
\multirow{3}{*}{\textbf{Methods}} &
\multicolumn{3}{c}{nDCG@20\vspace{0pt}} &
\multicolumn{3}{c}{nDCG@5\vspace{0pt}} &
\multicolumn{3}{c}{Precision@1\vspace{0pt}}\\
\cmidrule(lr){2-4} \cmidrule(lr){5-7} \cmidrule(lr){8-10}
& 
{Query} & {+ User Profile} & + Profile + Entities &
{Query} & {+ User Profile} & + Profile + Entities &
{Query} & {+ User Profile} & + Profile + Entities 
\\
\midrule
\Tstrut 
LM & 0.783 & \textbf{0.797} & 0.772 & 0.557 & \textbf{0.576} & 0.525 & 0.765 & \textbf{0.817} & 0.687\\
LM+WV & \textbf{0.782} & 0.781 & 0.776 & \textbf{0.558} & 0.548 & 0.541 & \textbf{0.748} & 0.722 & 0.704\\
BM25 & 0.781 & \textbf{0.799} & 0.776 & 0.555 & \textbf{0.579} & 0.536 & 0.774 & \textbf{0.809} & 0.696\\
DRMM  & 0.752 & \textbf{0.782} & - & 0.493 &\textbf{0.544} & - & 0.644 & \textbf{0.730} & - \\
PACRR & 0.766 &\textbf{0.792} & - & 0.53 & \textbf{0.572} & - & 0.67 & \textbf{0.809} & - \\
Commercial SE & 0.757 & - & - & 0.505 & - & - & 0.652 & - & -\\
\bottomrule
\end{tabular}
\caption{Results for different rankers when using the queries, user profiles, and user profiles enhanced by entity descriptions.}
\label{tab:results}

\end{table*}

\vspace{-0.1cm}
\section{Experiments}
\noindent{\bf Research Questions}\\
Our experiments 
target
the following
research questions.
\squishlist
\item[RQ1:] To what extent 
can sparse user profiles improve rankings towards individual interests? We compare non-personalized and personalized versions of all ranking methods 
of Section \ref{sec:method}.
\item[RQ2:] Do entity descriptions improve the ranking? 
We investigate ranking variants with and without 
entity models
(except for neural methods
which cannot handle long text as query input).
\item[RQ3:] Do word embeddings improve the ranking by 
semantic similarities between terms? 
We examine this by running the language-model-based ranker with and without
embeddings.
\squishend

\vspace{0.1cm}
\noindent{\bf Data}\\
We conducted an Amazon MTurk study in which we recruited 33 people for a two-stage task.
In the first stage, users create sparse profiles via questionnaires as explained in Section \ref{sec:user_profile}. In the second stage, for search-result evaluation, we asked the same workers for their judgements of {\em interestingness} on results of a small set of self-selected queries
(which they deemed of personal interest).
Queries were medium-grained 
gathered by crowdsourcing.
Examples from a total set of 50 queries are:
african books, 
greek mythology,
historical fiction,
memoirs and autobiography,
novels made into movies, 
scandinavian suspense,
sword and sorcery, 
time travel, 
true crime,
and
victorian society.

The judgements were graded \textit{"not interesting" (0)}, \textit{"interesting" (1)}, \textit{"very interesting" (2)} or 
\textit{"don't know" (discarded)}. 
To ensure 
faithful judgements,
we required a justification sentence for each judgement; examples 
are shown in Figure \ref{fig:user_judgements}.
Our pool of results contained 50 queries and 100 results to be judged per query. As it was not feasible to evaluate all 100 results for each user
and query,
we randomly selected 20 answers per query for per-user assessment.
We did not choose the top 20 results from a baseline
ranking (or pool rankings) 
in order to reduce bias towards (global) popularity as a ranking criterion.
We obtained judgements for 
115 user-query pairs by 33 users with 47 distinct queries -- 
2163 judged items in total. 

\vspace{0.1cm}
\noindent{\bf Experimental Setup}\\To investigate the utility of sparse user profiles for our setting, 
we evaluated a range of retrieval methods both with and without the user profiles.
The methods' performance was evaluated using normalized Discounted Cumulative Gain (nDCG@5 and nDCG@20) and Precision@1 (P@1) applied to condensed lists, with all unjudged results filtered out.
This follows \cite{sakai2007alternatives}, which argues that this approach to handling partial judgements is preferable to other metrics.
To compute precision, 
a result was considered good when deemed ``very interesting'' or ``interesting'' by the respective judge. 

Due to the limited size of our dataset
(despite considerable
spending on MTurk), we minimized the tuning of hyper-parameters and picked reasonable defaults where possible.
With the neural models, we used ten-fold cross-validation (with each of the ten folds containing unique queries):
eight folds for training
and the remaining two for validation and testing.
These models were trained 
 with a softmax loss over pairs of documents.
Due to the substantial increase in query length, we do not consider 
using entity descriptions with neural models.
We used the following hyper-parameters:
\squishlist
\item $\lambda$ determines the relative influence of
query and user models.
To test extreme cases,
we either set $\lambda=0$ or $\lambda=1$. 
\item $\mu$, the Dirichlet prior parameter for smoothing, was set to the average document length for each query.
\item N-gram order in the language model approach is set to 1.
\item When incorporating embedding similarity into the language model, we discarded terms with a similarity below $T=0.5$.
\item $\theta_C$ is the background model,
for which we use ClueWeb09.
\item We set $k_1=1.5$ and $b=0.75$ with BM25, after determining that a grid search on our folds could not improve results.
\item DRMM used an IDF gate with LCH-normalized histograms.
\item PACRR used unigrams through trigrams, 32 filters, a $k$-max of 2, and a combination layer of size 32.
\squishend

\vspace{-0.2cm}
\section{Results}

The results are shown in Table \ref{tab:results}
(LM denotes the language model method, LM+WV additionally
uses word2vec embeddings).
As a reference point, we include the quality of the original ranking obtained
from a commercial search engine.

Regarding RQ1, we observe that
incorporating sparse user profiles consistently improves results across methods and across metrics, with the exception of LM+WV.
We performed a paired t-test over all samples of 
non-personalized vs. personalized rankings; the p-values for
the hypothesis that personalization is beneficial were below 0.01
for all metrics. 

Regarding RQ2 and RQ3, neither word embeddings nor
entity descriptions helped to improve rankings, though.
Their performance even dropped below the non-personalized
baselines.
We believe that the word2vec embeddings are too broad and
diluted the focus of our queries.
This would call for user-specific embeddings, but it is an open
issue on how to obtain these.
The negative impact of entity descriptions 
is due to the 
breadth of entities including locations, movies, books and general concepts. %
Analyzing incorrectly high-ranked results indicated that
cues from the out-of-scope entity descriptions can be misleading due to the ambiguity of words. This calls for extending our model to consider other metrics such as entity specificity (i.e. selectively using entities and descriptions).

\vspace{0.1cm}
\noindent{\bf Ablation Study:}
We performed an ablation study with further restriction of
user profiles, to investigate how minimal the PKG 
could be while still being beneficial.
We evaluated two variants: 1) omitting the book-related fields
from questionnaires (but keeping fields on movies and music),
and 2) keeping solely the demographic attributes
and the hobbies field.
The first restriction led a notable loss in nDCG and precision,
but still gave decent quality, whereas the variant with minimal
profiles degraded.
For example, when using the BM25 ranker with 
80\% nDCG@20  and 81\%
precision@1 with full profiles,
leaving out the book fields gave 79\% nDCG@20 
and 77\% precision@1.
Capturing the user's interests
and tastes is crucial, but does not have to be 
domain-specific, like books.

\vspace{-0.15cm}
\section{Conclusion}
\vspace{-0.05cm}
Given the limited scope and scale of our study,
the results and findings are preliminary.
Nevertheless, they indicate that even sparse user profiles
have potential to improve ranking quality through personalization.
Our work in progress involves comparisons to richer profiles
obtained by gathering chat-like user-to-user conversations.
The overriding goal is to understand upper bounds of personalized 
quality is and gain insight on how well 
these can be approximated with different
extents of user profiling.
\begin{acks}
This research was supported by the ERC Synergy Grant 610150 (imPACT).
\end{acks}
\bibliographystyle{ACM-Reference-Format}
\bibliography{pse}

\end{document}